\documentclass[a4paper,11pt]{article}
\usepackage{pos}
\usepackage{wrapfig}

\title{Observation of Cosmic Ray Anisotropy with Nine Years of IceCube Data}
 \ShortTitle{Nine Years of Cosmic Ray Anisotropy}

\author{The IceCube Collaboration \\{\normalsize \normalfont(a complete list of authors can be found at the end of the proceedings)}}

\emailAdd{mcnally\_ft@mercer.edu}

\abstract{The IceCube Observatory has collected over 577 billion cosmic-ray induced muon events in its final configuration from May 2011 to May 2020. We used this data set to provide an unprecedented statistically accurate map of the cosmic ray arrival direction distribution in the TeV-PeV energy range scale in the Southern Hemisphere. Such an increase in event statistics makes it possible to extend the sensitivity to anisotropies at higher cosmic ray energies and smaller angular scales. It will also facilitate a more detailed assessment of the observatory stability over both short- and long-time scales. This will enable us to study the time variability of the cosmic ray anisotropy on a yearly-base and over the entire data sample period covering most of the solar cycle 24. We present the preliminary results from the study with the extended event sample.

\vspace{4mm}
{\bfseries Corresponding authors:}
Frank McNally$^{1*}$, Rasha Abbasi$^{2}$, Paolo Desiati$^{3}$, Juan Carlos D\'iaz V\'elez$^{3}$, Timothy Aguado$^{2}$, Katherine Gruchot$^{2}$, Andrew Moy$^{2}$, Alexander Simmons$^{1}$, Andrew Thorpe$^{1}$, and Hannah Woodward$^{4}$\\
{$^{1}$ \itshape Mercer University, U.S.A.}\\
{$^{2}$ \itshape Loyola University Chicago, U.S.A.}\\
{$^{3}$ \itshape University of Wisconsin - Madison, U.S.A.}\\
{$^{4}$ \itshape University of Virginia, U.S.A. (UW-Madison REU student 2020)}\\[4mm]
$^*$ Presenter

\FullConference{37$^{\rm{th}}$ International Cosmic Ray Conference (ICRC 2021)\\
		July 12th -- 23rd, 2021\\
		Online -- Berlin, Germany}

}

\begin{document}
\maketitle

\section{Introduction}

With the recent discovery of galactic gamma-ray Pevatrons~\cite{pevatron-tibet, pevatron-lhaaso} we are getting closer to the identification of the cosmic ray sources contributing to observations up to the knee. Traditionally, the measurement of cosmic ray energy and mass composition has been utilized by astrophysicists to infer global properties about their sources and their long propagation journey across the galactic magnetized medium. Now that their sources may have been identified, it is possible to improve our knowledge on how cosmic rays diffuse in the interstellar medium.
Anisotropy --- a property long observed in the arrival directions of cosmic rays --- may provide additional hints linking cosmic rays to their sources of acceleration and, more likely, to their propagation through the inhomogeneous interstellar medium.
On the one hand, ultra-high energy cosmic rays, whose direction is affected very little by astrophysical magnetized plasmas, can be used to identify their faraway extra-galactic sources~\citep{auger_2017, TA_2018}. On the other hand, the arrival direction distribution of lower energy cosmic rays, of galactic origin, may be used to unfold the properties of the interstellar medium through which they propagated~\cite{smowmass2021}.

Ground-based experiments have observed the cosmic rays' anisotropy since the seventies, but in the last two decades a new generation of observatories has provided unprecedented and detailed observations over a wide energy range (TeV-PeV)~\citep{nagashima_1998, hall_1999, amenomori_2005, amenomori_2006, amenomori_2007, guillian_2007, abdo_2008, abdo_2009, aglietta_2009, zhang_2009, munakata_2010, amenomori_2011, dejong_2011, shuwang_2011, bartoli_2013, abeysekara_2014, bartoli_2015, amenomori_2017, bartoli_2018, abeysekara_2018, abbasi_2010, abbasi_2011, abbasi_2012, aartsen_2013, aartsen_2016, bourbeau_2017, jcdv_2017, fermi_2019}. The anisotropy is small (of order of 10$^{-3}$ in relative intensity) and it changes with energy as if a transition between two distinct causes occurs around 100 TeV. Observations also show that the anisotropy has a complex angular structure, spanning from the large-scale dipole and quadrupole (thought to be associated with scattering off magnetic turbulence leading to pitch angle diffusion~\cite{giacinti_kirk_2017}) down to smaller scales with relative amplitude smaller than 10$^{-4}$ (thought to be linked to non-diffusive processes within a scattering mean path~\citep{giacinti_sigl_2012, lb_2017}). While new, large, ground-based experiments under construction (LHAASO~\citep{disciascio_2016}) and under design (SWGO~\citep{swgo_whitepaper, swgo_astro2020}, IceCube-Gen2~\cite{icecube-gen2}) will provide a leap in our anisotropy observations, existing experiments are still improving the quality of their data and analyses.

This work presents an update of cosmic ray anisotropy measurements in the energy range of 10 TeV to a few PeV, observed by the IceCube Observatory using the 577 billion events collected from 2011 to 2020. The unprecedented stability of the data sample, acquired by using the same experimental configuration over time and by improving analysis techniques, makes it possible for the first time to probe into time variabilities of the anisotropy with minimal systematic uncertainties.

\section{Data Analysis}

This work is designed to serve as an update to previous anisotropy studies completed in IceCube~\citep{abbasi_2010, abbasi_2011, abbasi_2012, aartsen_2016, bourbeau_2017}. We will therefore first highlight changes and improvements unique to this study before providing a brief overview of the overall method. 

The dataset for this work consists of 577 billion cosmic ray events, collected over nine years by the in-ice component of IceCube. In addition to being >80\% larger than the previous six-year publication~\citep{aartsen_2016}, these events were all collected in the completed 86-string configuration, allowing for unprecedented statistics and stability in systematic checks. The 9-year time frame brings the observation period much closer to covering an 11-year solar cycle. A consistent detector configuration also means we can use a single simulation dataset in the creation and study of energy-dependent skymaps. In this study, we use improved simulations featuring post-LHC CORSIKA SIBYLL 2.3c and a factor of 10 increase in statistics. The simulations include an improved description of the detector and of the optical properties of the instrumented ice. The Monte Carlo data is weighted to a Gaisser-H4a~\citep{gaisser_2012, gaisser_2013} composition model.

\begin{wrapfigure}{R}{0.5\textwidth}
  \begin{center}
    \includegraphics[width=0.48\textwidth]{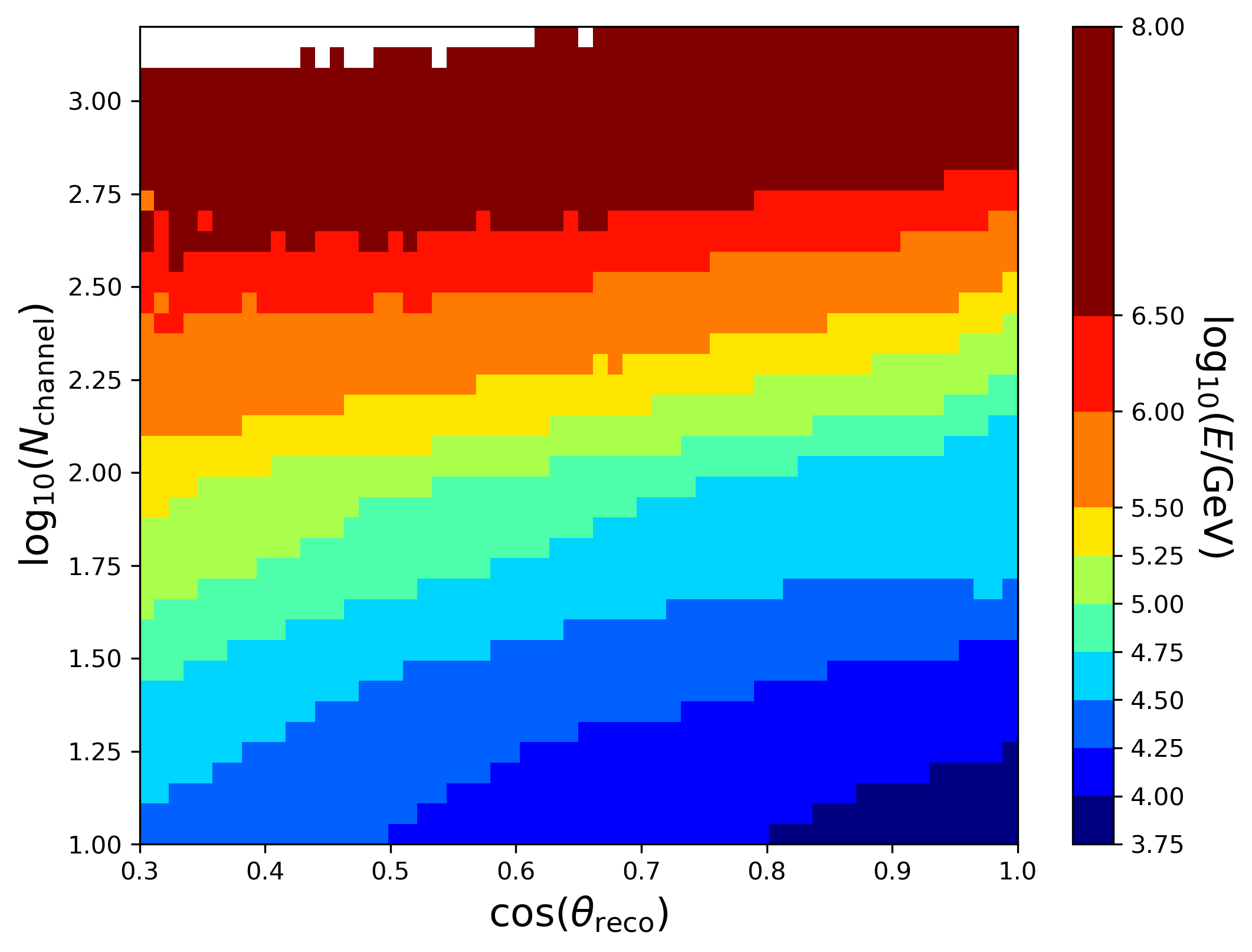}
  \end{center}
  \caption{Median true energy of cosmic ray primary particles (from simulation) as a function of reconstructed zenith ($\theta_\mathrm{reco}$) and the number of DOMs triggered ($N_\mathrm{channel}$).}
  \label{fig:med_energy}
\end{wrapfigure}

Aside from the aforementioned improvements, the analysis method was largely kept consistent with previous analyses~\citep{abbasi_2010, abbasi_2011, abbasi_2012, aartsen_2013, aartsen_2016, bourbeau_2017, jcdv_2017} to allow for direct comparison of the results. The maps that follow rely on the comparison of detector data to a background map that simulates the detector response to an isotropic sky. The background map itself is not isotropic, but must account for real-world factors, including detector response as a function of atmospheric depth, and preferred azimuthal acceptances due to detector configuration. To create the background maps, we use the time-scrambling method detailed in~\citep{abbasi_2011}. In short, we store the arrival times of all events within a time-scrambling window. Then, for each event, we create 20 background events, each with a weight of 1/20 and a time randomly selected from the time-scrambling window. By using local arrival coordinates and actual event times, the background map reflects the local angular and temporal event distribution seen by the detector. However, by choosing times from within a 24-hour time-scrambling window, the sidereal locations of the background events are randomly distributed in right ascension, producing a uniform distribution folded with the detector response.

All maps are created using HEALPix\footnote{https://healpix.jpl.nasa.gov} software, which bins the sky into equal area bins aligned along declination bands. Setting the parameter $N_\mathrm{side} = 64$ produces bins with an angular size of $(0.84^\circ)^2$. The relative intensity $\delta I$ is then calculated as a fractional deviation of the data $N$ from background $\langle N \rangle$, using $\delta I = (N - \langle N \rangle)/ \langle N \rangle$ on a pixel-by-pixel basis. Significance values are pre-trial, and calculated according to Li \& Ma~\citep{lima_1983}. The maps are smoothed by assigning to each pixel the combined value of all pixels within a given angular radius, a process known as ``top-hat" smoothing. For maps created from the full dataset, a $5^\circ$ smoothing radius is used, roughly corresponding to the $\ell$-value at which the power spectrum for IceCube is consistent with noise. For maps split by energy, a $20^\circ$ smoothing radius is used. This value was previously selected in order to maximize statistical significance while preserving our ability to observe large-scale structural changes.

To split the events into energy bins, we consider two parameters correlated with primary energy: the number of Digital Optical Modules (DOMs) hit ($N_\mathrm{channel}$), and the reconstructed zenith angle ($\theta_\mathrm{reco}$). While $N_\mathrm{channel}$ is a simple counter that does not take into account track location or detector geometry, it should still, on average, increase with primary energy. The zenith-dependence of primary energy arises from the atmospheric depth the shower must penetrate before reaching the detector; lower-energy showers are filtered out at zenith angles closer to the horizon, their mean particle energy dropping below the detection threshold before the shower reaches the detector. It should be noted that, because IceCube is only sensitive to the muonic component of air showers, there are inherent limits on the energy resolution. As a result, while the energy bins are statistically independent, the energy distributions of adjacent energy bins have significant overlap. Using simulation, all events were binned in $\log_{10}(N_\mathrm{channel})$ and $\cos (\theta_\mathrm{reco})$. Figure~\ref{fig:med_energy} shows the resultant median energy value for each bin. After smoothing this table using splines (see method in~\citep{whitehorn_2013}), the $N_\mathrm{channel}$ and $\theta_\mathrm{reco}$ values of each event were used to place it in an energy bin.

\section{Results}

\begin{figure*}[ht]
  \centering
  \includegraphics[width=0.49\textwidth]{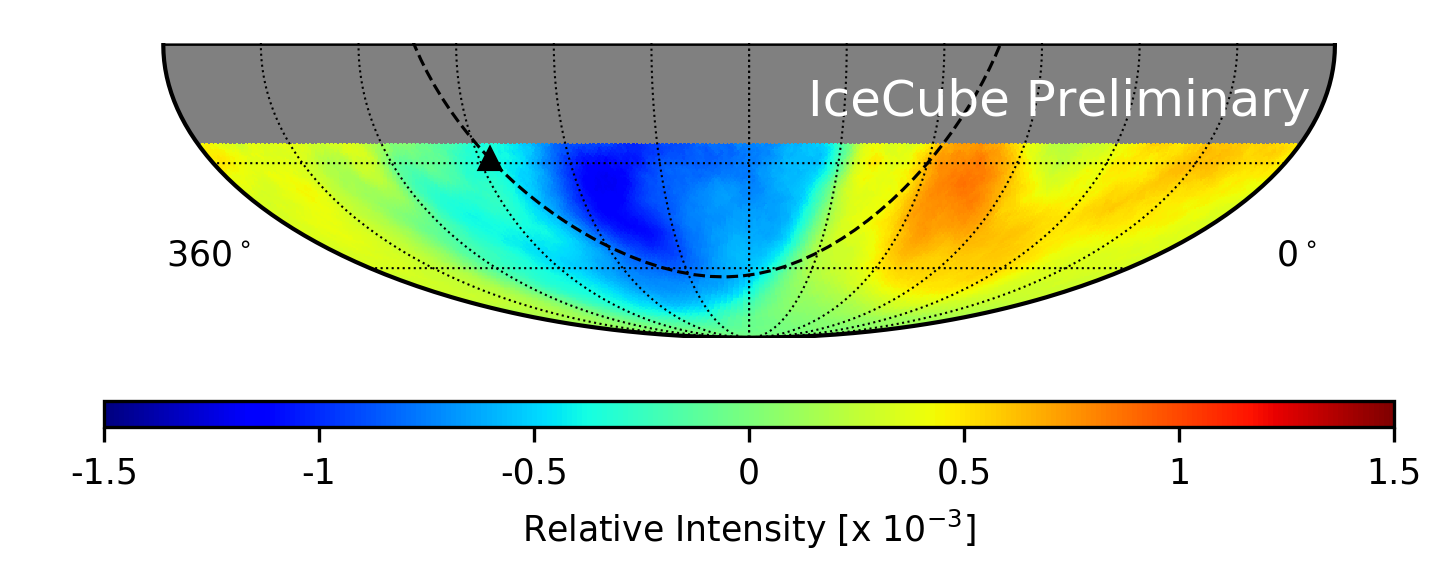}
  \includegraphics[width=0.49\textwidth]{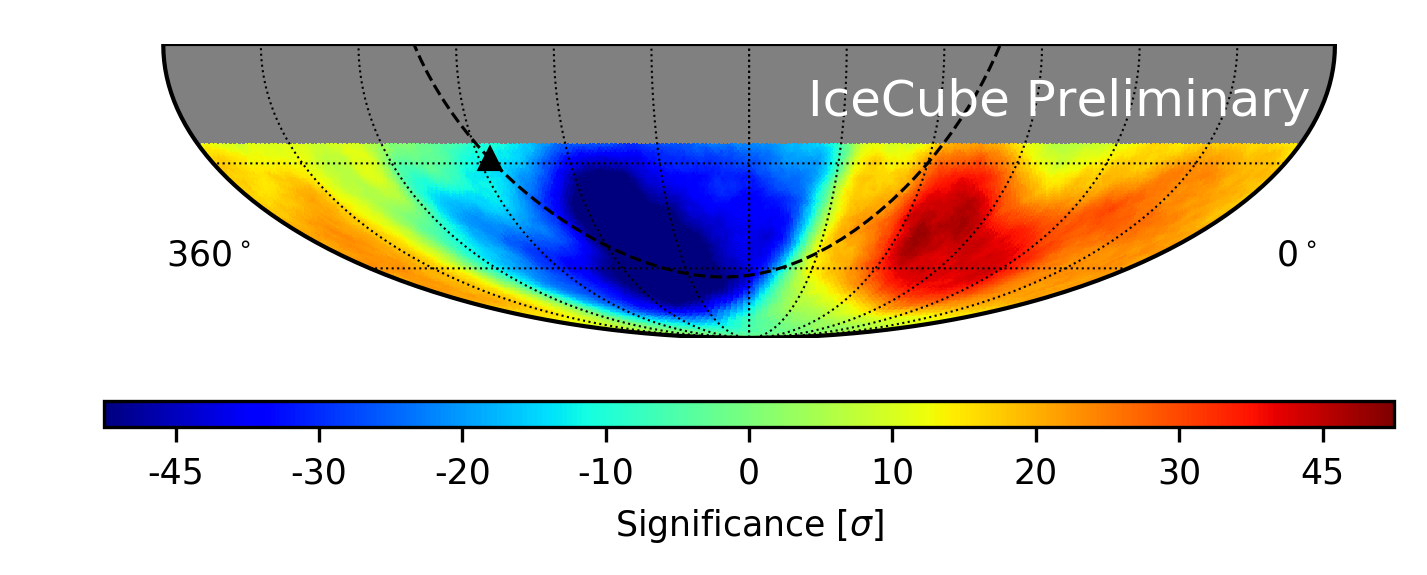}
  \includegraphics[width=0.49\textwidth]{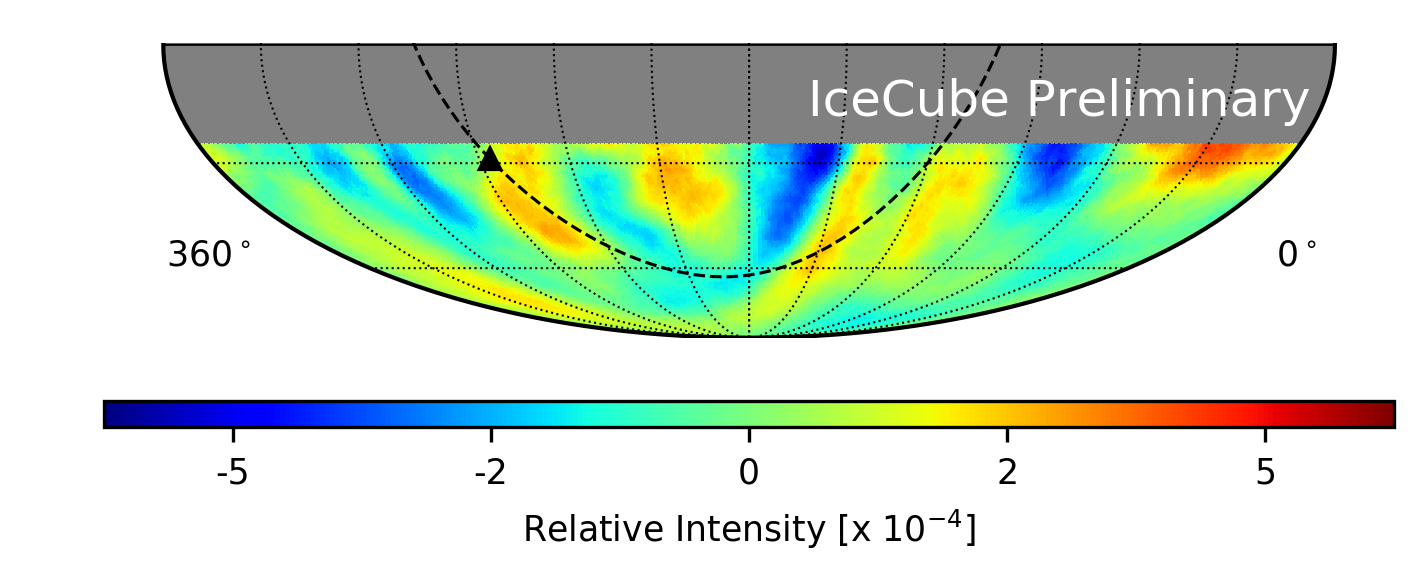}
  \includegraphics[width=0.49\textwidth]{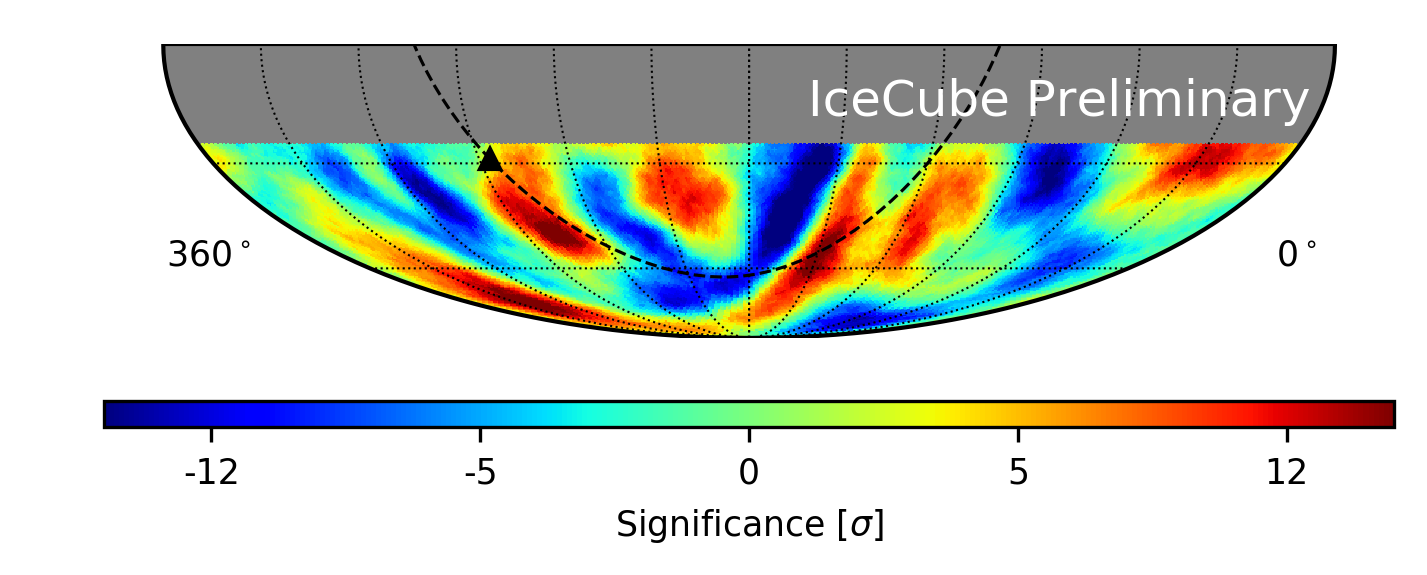}
  \caption{Relative intensity (\textit{left}) and significance (\textit{right}) maps, shown in equatorial coordinates for all data (\textit{top}) and with best-fit dipole and quadrupole terms subtracted (\textit{bottom}). The dashed line indicates the galactic plane, the triangle marks the galactic center.}
  \label{fig:largesmall}
\end{figure*}

A study of the sidereal anisotropy using all events is shown in Fig.~\ref{fig:largesmall}. These maps are consistent with previous studies --- at both large and small angular scales, we see the same features at the same locations. This study does yield higher pre-trial significance values, as expected given the increased sample size.

\begin{figure}[ht]
  \centering
  \includegraphics[width=\textwidth]{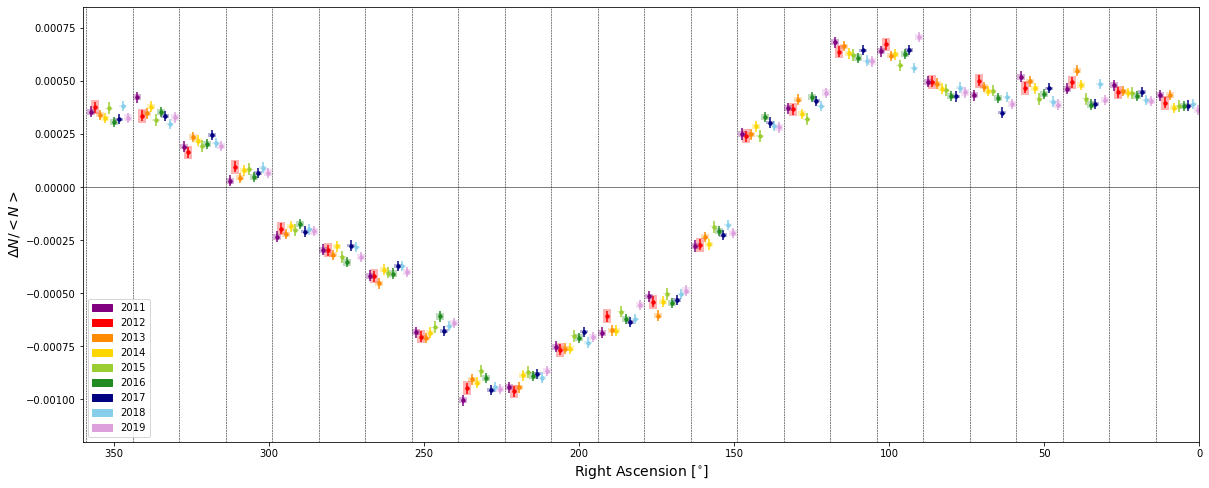}
  \caption{A one-dimensional projection of relative intensity binned in right ascension, split by detector year. The small, shaded regions around each point represent systematic uncertainties, calculated using the amplitude of the best-fit dipole to the anti-sidereal distribution for each year. The solid error bars are statistical.}
  \label{fig:sid_1d}
\end{figure}

To better visualize any potential time-dependence of the anisotropy, we can look at a one-dimensional projection of the data along right ascension. To produce Fig.~\ref{fig:sid_1d}, the data was split by calendar year, from May $15^\mathrm{th}$ of the listed year to May $14^\mathrm{th}$ of the following. The shift from detector to calendar years provides a better grasp on our systematic uncertainties. The length of detector years is somewhat arbitrary, with some being much longer than others. When a time period significantly greater or less than 365 days is considered, the Compton-Getting effect due to the Earth's motion around the Sun does not cancel out. In the frequency domain, we are concerned with the contamination of the 366-day sidereal year by sidebands from the 365-day solar year. To determine the magnitude of this effect, we consider a 364-day ``anti-sidereal" year~\citep{guillian_2007}. By shifting to calendar years, the influence of any solar signal on the sidereal frame is minimized, visible as a diminished signal in the anti-sidereal frame. As a result, the systematic error bars, which once were much larger than the statistical errors, are now effectively equivalent. With the newly reduced systematic uncertainties and extended observation time, there appear to be time-related shifts in several of the right ascension bins. Studies are currently underway to determine the significance of the apparent effect, incorporating a recently-collected tenth year of data.

\begin{figure}[ht]
  \centering
  \includegraphics[width=\textwidth]{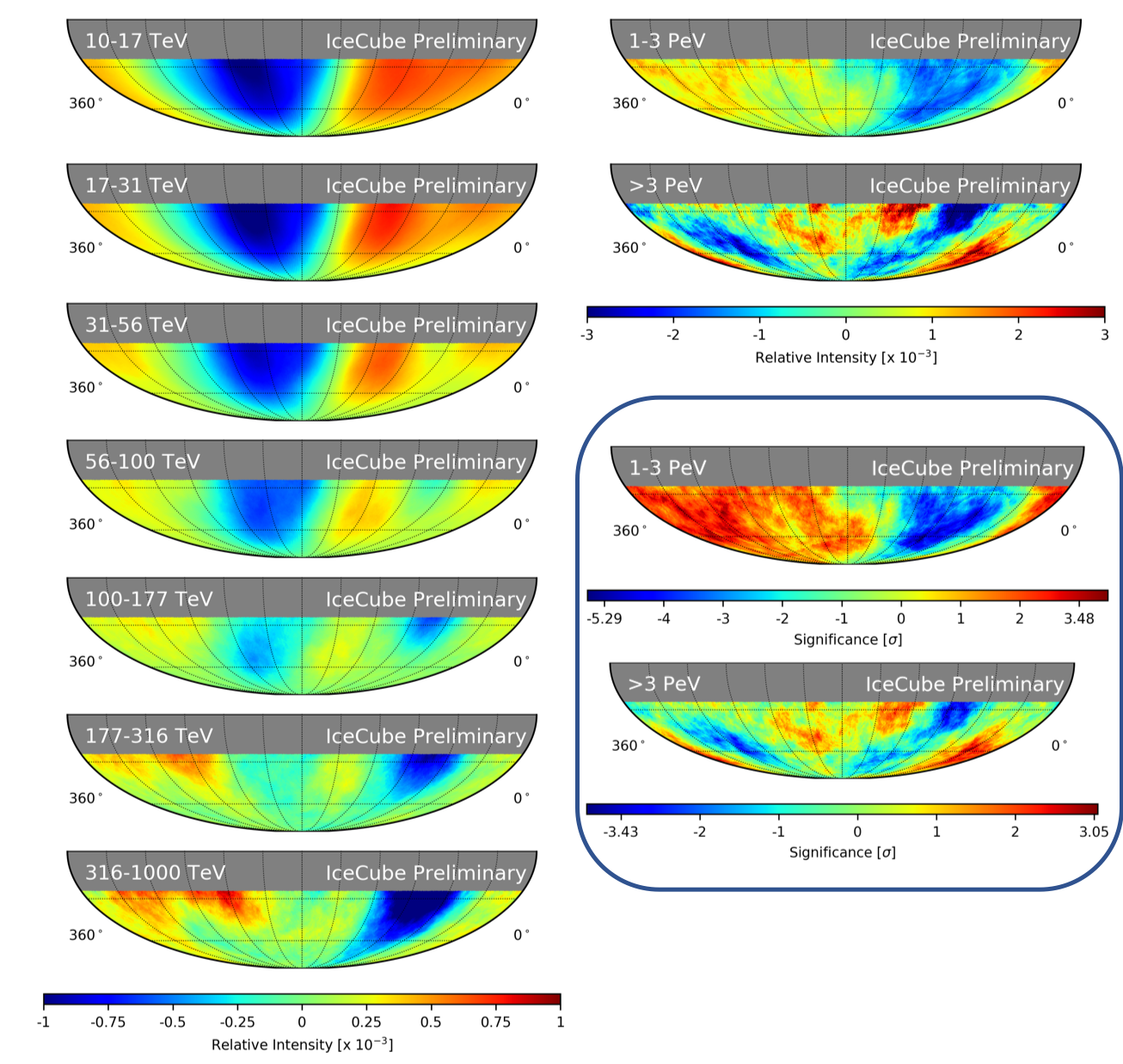}
  \caption{Relative intensity maps for each energy bin (\textit{left, top right}) featuring top-hat smoothing with a $20^\circ$ angular radius. Both the features present and the change in structure are consistent with previous work. The scale is changed at 1\,PeV to avoid saturation. Also shown are significance maps for the two highest energy bins (\textit{bottom right}).}
  \label{fig:ebins}
\end{figure}

The new simulation yields energy-dependent maps that are also consistent with previous results, as shown in Fig.~\ref{fig:ebins}. Large regions of excess and deficit ($+90^\circ$ and $+270^\circ$, respectively) fade in strength with increasing energy, replaced by a dominant deficit near a right ascension of $+90^\circ$ around 100\,TeV. While the locations and relative intensities of these structures are consistent with previous studies, the increased sample size pushes several of the features in the highest-energy map to $>3\sigma$ (see Fig.~\ref{fig:ebins}). The significance of the features present in the 1-3\,PeV map also changes; studies are underway to determine whether this is a time-dependent result, or an effect of the new simulation used to create our energy bins.

\section{Conclusions}

We present several improvements to previous IceCube analyses of the observed anisotropy in cosmic ray arrival direction. Nine years of data collection with the 86-string configuration of the in-ice component of IceCube provide an unprecedented sample size of 577 billion events. We correspondingly see large- and small-scale structures in the all-event sample at higher significance, with the morphology of these maps largely unchanged. A study of the energy-dependence yields a similar result; while the observed structures and transitions remain consistent, we can report features in our highest-energy map (>3\,PeV) at a pre-trial significance of over $3\sigma$. Changes to the energy maps are a function of both enhanced statistics and improvements to event simulation.

Use of a consistent detector configuration also enables the study of data by calendar year as opposed to detector year. Systematic uncertainties in the sidereal anisotropy arise from interference with the solar frame, an effect that should largely cancel over a calendar year. As a result, our systematic uncertainties in the time-dependent study of the sidereal anisotropy (Fig.~\ref{fig:sid_1d}) are significantly reduced. As IceCube approaches 11 years of data collection in its 86-string configuration, this work should enhance our ability to search for any effect of the 11-year solar cycle on the observed cosmic ray anisotropy.

\clearpage
\section*{Full Author List: IceCube Collaboration}




\scriptsize
\noindent
R. Abbasi$^{17}$,
M. Ackermann$^{59}$,
J. Adams$^{18}$,
J. A. Aguilar$^{12}$,
M. Ahlers$^{22}$,
M. Ahrens$^{50}$,
C. Alispach$^{28}$,
A. A. Alves Jr.$^{31}$,
N. M. Amin$^{42}$,
R. An$^{14}$
% ...
\\

\noindent
% ...
$^{12}$ Universit{\'e} Libre de Bruxelles, Science Faculty CP230, B-1050 Brussels, Belgium \\
% ...
$^{14}$ Department of Physics and Laboratory for Particle Physics and Cosmology, Harvard University, Cambridge, MA 02138, USA \\
% ...
$^{17}$ Department of Physics, Loyola University Chicago, Chicago, IL 60660, USA \\
$^{18}$ Dept. of Physics and Astronomy, University of Canterbury, Private Bag 4800, Christchurch, New Zealand \\
% ...
$^{22}$ Niels Bohr Institute, University of Copenhagen, DK-2100 Copenhagen, Denmark \\
% ...
$^{28}$ D{\'e}partement de physique nucl{\'e}aire et corpusculaire, Universit{\'e} de Gen{\`e}ve, CH-1211 Gen{\`e}ve, Switzerland \\
% ...
$^{31}$ Karlsruhe Institute of Technology, Institute for Astroparticle Physics, D-76021 Karlsruhe, Germany  \\
% ...
$^{42}$ Bartol Research Institute and Dept. of Physics and Astronomy, University of Delaware, Newark, DE 19716, USA \\
% ...
$^{50}$ Oskar Klein Centre and Dept. of Physics, Stockholm University, SE-10691 Stockholm, Sweden \\
% ...
$^{59}$ DESY, D-15738 Zeuthen, Germany \\
% ...

\subsection*{Acknowledgements}

\noindent
USA {\textendash} U.S. National Science Foundation-Office of Polar Programs,
U.S. National Science Foundation-Physics Division,
U.S. National Science Foundation-EPSCoR,
Wisconsin Alumni Research Foundation,
Center for High Throughput Computing (CHTC) at the University of Wisconsin{\textendash}Madison,
Open Science Grid (OSG),
Extreme Science and Engineering Discovery Environment (XSEDE),
Frontera computing project at the Texas Advanced Computing Center,
U.S. Department of Energy-National Energy Research Scientific Computing Center,
Particle astrophysics research computing center at the University of Maryland,
Institute for Cyber-Enabled Research at Michigan State University,
and Astroparticle physics computational facility at Marquette University;
Belgium {\textendash} Funds for Scientific Research (FRS-FNRS and FWO),
FWO Odysseus and Big Science programmes,
and Belgian Federal Science Policy Office (Belspo);
Germany {\textendash} Bundesministerium f{\"u}r Bildung und Forschung (BMBF),
Deutsche Forschungsgemeinschaft (DFG),
Helmholtz Alliance for Astroparticle Physics (HAP),
Initiative and Networking Fund of the Helmholtz Association,
Deutsches Elektronen Synchrotron (DESY),
and High Performance Computing cluster of the RWTH Aachen;
Sweden {\textendash} Swedish Research Council,
Swedish Polar Research Secretariat,
Swedish National Infrastructure for Computing (SNIC),
and Knut and Alice Wallenberg Foundation;
Australia {\textendash} Australian Research Council;
Canada {\textendash} Natural Sciences and Engineering Research Council of Canada,
Calcul Qu{\'e}bec, Compute Ontario, Canada Foundation for Innovation, WestGrid, and Compute Canada;
Denmark {\textendash} Villum Fonden and Carlsberg Foundation;
New Zealand {\textendash} Marsden Fund;
Japan {\textendash} Japan Society for Promotion of Science (JSPS)
and Institute for Global Prominent Research (IGPR) of Chiba University;
Korea {\textendash} National Research Foundation of Korea (NRF);
Switzerland {\textendash} Swiss National Science Foundation (SNSF);
United Kingdom {\textendash} Department of Physics, University of Oxford.

%
%
%


\begin{thebibliography}{99}

\bibitem[Amenomori et al.(2021)]{pevatron-tibet}\textbf{Tibet AS$\gamma$} Collaboration, (2021), Nature Astronomy 5, 460
\bibitem[Cao et al.(2021)]{pevatron-lhaaso}\textbf{LHAASO} Collaboration, Cao, Z. et al., (2021), Nature 594, 33
\bibitem[Aab et al.(2017)]{auger_2017}\textbf{Pierre Auger} Collaboration, Aab, A. et al., (2017), Science 357, 1266
\bibitem[Abbasi et al.(2018)]{TA_2018}\textbf{TA} Collaboration, Abbasi, R.U. et al., (2018) Astrophys. J. Lett., 867, L27
\bibitem[Desiati et al.(2020)]{smowmass2021}Desiati, P., et al., (2021), Snowmass 2021 Letter of Intent, arXiv:2009.04883
\bibitem[Nagashima et al.(1988)]{nagashima_1998}Nagashima, et~al. (1998), J. of Geophys. Res. 1031, 17429
\bibitem[Hall et al.(1999)]{hall_1999}Hall, D.L. et al., (1999), J. of Geophys. Res. 104, 6737
\bibitem[Amenomori et al.(2005)]{amenomori_2005}\textbf{Tibet AS$\gamma$} Collaboration, Amenomori, M. et~al., (2005), Astrophys. J. Lett. 626, L29
\bibitem[Amenomori et al.(2006)]{amenomori_2006}\textbf{Tibet AS$\gamma$} Collaboration, Amenomori, M. et~al., (2006), Science, 314, 439
\bibitem[Amenomori et al.(2007)]{amenomori_2007}\textbf{Tibet AS$\gamma$} Collaboration, Amenomori, M. et~al., (2007), Proc. 30th ICRC, M\'erida, Mexico
\bibitem[Guillian et al.(2007)]{guillian_2007}\textbf{Super-Kamiokande} Collaboration, Guillian, G. et al., (2007), Phys. Rev. D 75, 062003
\bibitem[Abdo et al.(2008)]{abdo_2008}\textbf{Milagro} Collaboration, Abdo, A.A. et al., (2008), Phys. Rev. Lett. 101, 221 101
\bibitem[Abdo et al.(2009)]{abdo_2009}\textbf{Milagro} Collaboration, Abdo, A.A. et al., (2009), Astrophys. J. 698, 2121
\bibitem[Aglietta et al.(2009)]{aglietta_2009}\textbf{EAS-TOP} Collaboration, Aglietta, M. et~al., (2009), Astrophys. J. 692, L130
\bibitem[Zhang et al.(2009)]{zhang_2009}\textbf{ARGO-YBJ} Collaboration, Zhang, J.L., (2009), Proc. 31st ICRC, \L\'od\'z, Poland
\bibitem[Munakata et al.(2010)]{munakata_2010}Munakata, K. et~al. (2010), Astrophys. J. 712, 1100
\bibitem[Amenomori et al.(2011)]{amenomori_2011}\textbf{Tibet AS$\gamma$} Collaboration, Amenomori, M. et~al., (2011), Proc. 32nd ICRC, Beijing China
\bibitem[de Jong et al.(2011)]{dejong_2011}\textbf{MINOS} Collaboration, de Jong, J. et al., (2011), Proc. 32nd ICRC, Beijing, China
\bibitem[Shuwang et al.(2011)]{shuwang_2011}\textbf{ARGO-YBJ} Collaboration, Shuwang, C., et. al. (2011), Proc. 32nd ICRC, Beijing China
\bibitem[Bartoli et al.(2013)]{bartoli_2013}\textbf{ARGO-YBJ} Collaboration, Bartoli, B., et~al., (2013) Phys. Rev. D 88-8, 082001
\bibitem[Abeysekara et al.(2014)]{abeysekara_2014}\textbf{HAWC} Collaboration, Abeysekara, A.U. et al., (2014), Astrophys. J. 796, 108
\bibitem[Bartoli et al.(2015)]{bartoli_2015}\textbf{ARGO-YBJ} Collaboration, Bartoli, B., et~al., (2015), Astrophys. J. 809, 90
\bibitem[Amenomori et al.(2017)]{amenomori_2017}\textbf{Tibet AS$\gamma$} Collaboration, Amenomori, M. et~al., (2017), Astrophys. J. 836, 153
\bibitem[Bartoli et al.(2018)]{bartoli_2018}\textbf{ARGO-YBJ} Collaboration, Bartoli, B., et~al., (2018), Astrophys. J., 861, 93
\bibitem[Abeysekara et al.(2018)]{abeysekara_2018}\textbf{HAWC} Collaboration, Abeysekara, A.U. et al. (2018), Astrophys. J., 865, 57
\bibitem[Abbasi et al.(2010)]{abbasi_2010}\textbf{IceCube} Collaboration, Abbasi, R. et al., (2010) Astrophys. J. 718, L194
\bibitem[Abbasi et al.(2011)]{abbasi_2011}\textbf{IceCube} Collaboration, Abbasi, R. et al., (2011), Astrophys. J. 740 16
\bibitem[Abbasi et al.(2012)]{abbasi_2012}\textbf{IceCube} Collaboration, Abbasi et al., (2012), Astrophys. J. 746, 33
\bibitem[Aartsen et al.(2013)]{aartsen_2013}\textbf{IceCube} Collaboration, Aartsen, M. et al., (2013), Astrophys. J. 765, 55
\bibitem[Aartsen et al.(2016)]{aartsen_2016}\textbf{IceCube} Collaboration, Aartsen, M. et al., (2016), Astrophys. J. 826, 220
\bibitem[Aartsen et al.(2017)]{bourbeau_2017}\textbf{IceCube} Collaboration, Aartsen M., et al., (2017), PoS(ICRC2017)474
\bibitem[HAWC \& IceCube Collaborations(2019)]{jcdv_2017} \textbf{HAWC \& IceCube} Collaboration, (2019), Astrophys. J., 871, 96
\bibitem[Ajello et al.(2019)]{fermi_2019}\textbf{Fermi-LAT} Collaboration, Ajello, M., et al., (2019), arXiv:1903.02905
\bibitem[Giacinti \& Kirk(2017)]{giacinti_kirk_2017}Giacinti, G. \& Kirk, J.G., (2017), Astrophys. J. 835, 258 
\bibitem[Giacinti \& Sigl(2012)]{giacinti_sigl_2012}Giacinti, G., \& Sigl, G. (2012) Phys. Rev. Lett. 109, 071101
\bibitem[L\'opez-Barquero et al.(2017)]{lb_2017} L\'opez-Barquero, V., et al. (2017), Astrophys. J., 842, 54
\bibitem[Di Sciascio et al.(2016)]{disciascio_2016}Di Sciascio, G. \& Iuppa, R., (2016), Nucl. and Part. Phys. Proc., 279, 166
\bibitem[Albert et al.(2019)]{swgo_whitepaper}Albert A., et al., (2019) arXiv:1902.08429
\bibitem[Abreu et al.(2019)]{swgo_astro2020}Abreu P., et al., (2019) Astro2020 APC White Paper, arXiv:1907.07737
\bibitem[Aartsen et al.(2021)]{icecube-gen2}\textbf{IceCube} Collaboration, Aartsen, M., et al., (2021), J. Phys. G 6, 060501
\bibitem[Gaisser (2012)]{gaisser_2012}Gaisser, T.,(2012), Astropart. Phys., 35, 12
\bibitem[Gaisser et al.(2013)]{gaisser_2013}Gaisser, T., et al.,(2013), Front. Phys. (\textit{Beijing}), 8, 748-758
\bibitem[Li \& Ma (1983)]{lima_1983}Li, T.-P., \& Ma, Y.-Q., (1983), Astrophys. J., 272, 317
\bibitem[Whitehorn et al.(2013)]{whitehorn_2013}Whitehorn, N., et al., (2013), Comput. Phys. Commun., 184, 2214-2220



\end{thebibliography}
\end{document}